\documentclass[twoside,twocolumn,9pt]{article}
\usepackage{extsizes}
\usepackage[super,sort&compress,comma]{natbib} 
\usepackage[left=1.5cm, right=1.5cm, top=1.785cm, bottom=2.0cm]{geometry}
\usepackage{balance}
\usepackage{mathptmx}
\usepackage{sectsty}
\usepackage{graphicx} 
\usepackage{lastpage}
\usepackage[format=plain,justification=justified,singlelinecheck=false,font={stretch=1.125,small,sf},labelfont=bf,labelsep=space]{caption}
\usepackage{float}
\usepackage{fancyhdr}
\usepackage{fnpos}
\usepackage[english]{babel}
\addto{\captionsenglish}{%
  
}
\usepackage{array}
\usepackage{droidsans}
\usepackage{charter}
\usepackage[T1]{fontenc}
\usepackage[usenames,dvipsnames]{xcolor}
\usepackage{setspace}
\usepackage[compact]{titlesec}
\usepackage{hyperref}
\usepackage{epstopdf}%This line makes .eps figures into .pdf - please comment out if not required.

\definecolor{cream}{RGB}{222,217,201}

\begin{document}

\pagestyle{fancy}
\thispagestyle{plain}
\fancypagestyle{plain}{
%%%HEADER%%%
\renewcommand{\headrulewidth}{0pt}
}
%%%END OF HEADER%%%

%%%PAGE SETUP - Please do not change any commands within this section%%%
\makeFNbottom
\makeatletter
\renewcommand\LARGE{\@setfontsize\LARGE{15pt}{17}}
\renewcommand\Large{\@setfontsize\Large{12pt}{14}}
\renewcommand\large{\@setfontsize\large{10pt}{12}}
\renewcommand\footnotesize{\@setfontsize\footnotesize{7pt}{10}}
\makeatother

\renewcommand{\thefootnote}{\fnsymbol{footnote}}
\renewcommand\footnoterule{\vspace*{1pt}% 
\color{cream}\hrule width 3.5in height 0.4pt \color{black}\vspace*{5pt}} 
\setcounter{secnumdepth}{5}

\makeatletter 
\renewcommand\@biblabel[1]{#1}            
\renewcommand\@makefntext[1]% 
{\noindent\makebox[0pt][r]{\@thefnmark\,}#1}
\makeatother 
\renewcommand{\figurename}{\small{Fig.}~}
\sectionfont{\sffamily\Large}
\subsectionfont{\normalsize}
\subsubsectionfont{\bf}
\setstretch{1.125} %In particular, please do not alter this line.
\setlength{\skip\footins}{0.8cm}
\setlength{\footnotesep}{0.25cm}
\setlength{\jot}{10pt}
\titlespacing*{\section}{0pt}{4pt}{4pt}
\titlespacing*{\subsection}{0pt}{15pt}{1pt}
%%%END OF PAGE SETUP%%%

%%%FOOTER%%%
\fancyfoot{}
\fancyfoot[LO,RE]{\vspace{-7.1pt}\includegraphics[height=9pt]{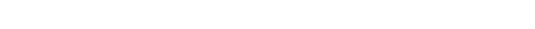}}
\fancyfoot[CO]{\vspace{-7.1pt}\hspace{13.2cm}\includegraphics{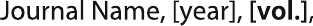}}
\fancyfoot[CE]{\vspace{-7.2pt}\hspace{-14.2cm}\includegraphics{RF.pdf}}
\fancyfoot[RO]{\footnotesize{\sffamily{1--\pageref{LastPage} ~\textbar  \hspace{2pt}\thepage}}}
\fancyfoot[LE]{\footnotesize{\sffamily{\thepage~\textbar\hspace{3.45cm} 1--\pageref{LastPage}}}}
\fancyhead{}
\renewcommand{\headrulewidth}{0pt} 
\renewcommand{\footrulewidth}{0pt}
\setlength{\arrayrulewidth}{1pt}
\setlength{\columnsep}{6.5mm}
\setlength\bibsep{1pt}
%%%END OF FOOTER%%%

%%%FIGURE SETUP - please do not change any commands within this section%%%
\makeatletter 
\newlength{\figrulesep} 
\setlength{\figrulesep}{0.5\textfloatsep} 

\newcommand{\topfigrule}{\vspace*{-1pt}% 
\noindent{\color{cream}\rule[-\figrulesep]{\columnwidth}{1.5pt}} }

\newcommand{\botfigrule}{\vspace*{-2pt}% 
\noindent{\color{cream}\rule[\figrulesep]{\columnwidth}{1.5pt}} }

\newcommand{\dblfigrule}{\vspace*{-1pt}% 
\noindent{\color{cream}\rule[-\figrulesep]{\textwidth}{1.5pt}} }

\makeatother
%%%END OF FIGURE SETUP%%%

%%%TITLE, AUTHORS AND ABSTRACT%%%
\twocolumn[
  \begin{@twocolumnfalse}
{%\includegraphics[height=30pt]{SM}\hfill\raisebox{0pt}[0pt][0pt]{\includegraphics[height=55pt]{RSC_LOGO_CMYK.pdf}}\\[1ex]
}\par
\vspace{1em}
\sffamily
\begin{tabular}{m{4.5cm} p{13.5cm} }

\includegraphics{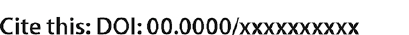} & \noindent\LARGE{\textbf{Self-Diffusion Scalings in Dense Granular Flows}} \\%Article title goes here instead of the text "This is the title"
\vspace{0.3cm} & \vspace{0.3cm} \\

% & \noindent\large{Full Name,$^{\ast}$\textit{$^{a}$} Full Name,\textit{$^{b\ddag}$} and Full Name\textit{$^{a}$}} \\%Author names go here instead of "Full name", etc.
 & \noindent\large{Riccardo Artoni,$^{\ast}$\textit{$^{a}$} Michele Larcher,\textit{$^{b}$} James T. Jenkins,\textit{$^{c}$} and Patrick Richard\textit{$^{a}$}} \\%Author names go here instead of "Full name", etc.
 
\includegraphics{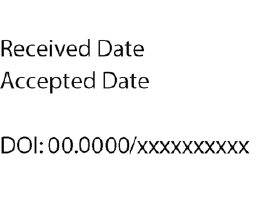} & \noindent\normalsize{We report on measurements of self-diffusion coefficients in discrete numerical simulations of steady, homogeneous, collisional shearing flows of nearly identical, frictional, inelastic spheres. We focus on a range of relatively high {solid} volume fractions that are important in those terrestrial gravitational shearing flows that are dominated by collisional interactions. Diffusion over this range of {solid} fraction has not been well characterized in previous studies. {We first compare the measured values with an empirical scaling based on shear rate previously proposed in the literature, and highlight the presence of anisotropy and the solid fraction dependence. We then compare the numerical measurements} with those predicted by the kinetic theory for shearing flows of inelastic spheres and offer an explanation for why the measured and predicted values differ.} \\%The abstrast goes here instead of the text "The abstract should be..."

\end{tabular}

 \end{@twocolumnfalse} \vspace{0.6cm}

  ]
%%%END OF TITLE, AUTHORS AND ABSTRACT%%%

%%%FONT SETUP - please do not change any commands within this section
\renewcommand*\rmdefault{bch}\normalfont\upshape
\rmfamily
\section*{}
\vspace{-1cm}

%%%FOOTNOTES%%%
\footnotetext{\textit{$^{a}$~MAST-GPEM, Univ Gustave Eiffel, IFSTTAR, F-44344 Bouguenais, France; E-mail: riccardo.artoni@univ-eiffel.fr}}
\footnotetext{\textit{$^{b}$~Free University of Bozen-Bolzano, I-39100 Bozen-Bolzano, Italy.}}
\footnotetext{\textit{$^{c}$~Cornell University, Ithaca, NY 14053, USA.}}

%%%END OF FOOTNOTES%%%

%%%MAIN TEXT%%%%
\section{Introduction}
Collisions between spheres in a dense granular shearing flow induce velocity fluctuations of the grains that drive the diffusion of particles in a fashion that is analogous to the thermal diffusion in a dense gas of elastic molecules, or the diffusion induced by eddies in the turbulent flow of a fluid. Here, we report on measurements of the components of  self-diffusion parallel and perpendicular to the flow done in discrete numerical simulations of inelastic spheres in a dense shearing flow. The components are determined by measuring the average squared displacement of spheres as a function of time.

Diffusion in granular shearing flows is important to mixing and segregation and has been studied experimentally in various flow geometries:  granular shear cells \cite{scott76,buggisch89}; vertical channels \cite{hunt92,natarajan95};  inclined chutes \cite{savage88,zik91, tripathi11}; vibrationally excited systems \cite{wildman06}; free-surface flows \cite{drahun83}; rotating tumblers \cite{alonso91,felix04,jain05a,jain05b}; and rotating tubes \cite{metcalfe96,khan05}. Savage \& Dai \cite{savage93} and Thornton, et al. \cite{thornton12} have carried out studies of segregation in discrete numerical simulations; and phenomenological theories exist, such as those described by Gray \& Ancey \cite{gray11} and Fan \& Hill \cite{fan11}, that produce plausible predictions of species’ concentrations and mixture velocity for appropriate choices of parameters. \citet{taberlet06} studied the spreading of a granular pulse in numerical simulations of bidisperse mixtures in a rotating drum. All of these studies are interpreted in terms of the mechanism of diffusion. {The ability of kinetic theory to predict diffusion properties in 2D granular systems sustained by an air table has also been studied extensively in~\cite{oger96} for a wide range of surface fractions .}

Direct experimental and numerical measurements of components of the tensor of self-diffusion exist in regimes of flow somewhat different from that considered here. Campbell \cite{campbell97} carried out such measurements in a sheared system of spheres that interacted through frictional, inelastic collisions over a range of restitution coefficients from 0.4 to 1.0 and {solid fractions} from 0.0001 to 0.5. Macaulay \& Rognon \cite{macaulay19} investigated the effect of inter-granular cohesive forces on the properties of self-diffusion in dense granular flows. Utter \& Behringer \cite{utter04} performed measurements in experiments on slow, rate-independent shearing of a dense aggregate of disks in a Couette cell. In two-dimensional numerical simulations of a similar system in a periodic cell, Radjai \& Roux \cite{radjai02} measured properties of the particle velocity fluctuations and the components of the self-diffusion. Because these systems involve rate-independent interactions, they differ from the system that we consider.

The measurements that we report are similar to Campbell's, in their common range of {solid} volume fraction; but different, in our focus on the range of volume fractions between 0.49 and 0.6. This is the range of {solid} fraction important in geophysical flows on Earth.
Our interest is in the values of the components of the tensor of self-diffusion, particularly over the range of {solid} fraction between 0.49, above which long range order might appear in equilibrated system of monosized elastic spheres 
 \cite{alder57}, and a {solid} fraction of about $\phi_c=0.587$, at which a collisional flow becomes impossible, when the coefficient of sliding friction is 0.5  \cite{berzi15}.

Previous studies have pointed out that self diffusion
coefficient scales simply, in dense systems, as $D=kd^2\dot\gamma$, where $\dot\gamma$ is the shear rate, $d$ the particle diameter, and $k$ approximately a
constant of order 0.05~\cite{utter04,fry19, cai19}.
This scaling is, however, empirical. In addition, in {such} dense systems, little attention has been given to the tensorial nature of self-diffusivity and to its dependence on solid fraction.  Therefore, our first goal is  to study the full self-diffusivity tensor over the range of dense {solid} fractions from 0.49 to 0.587. In addition, to our knowledge, the scaling based on shear rate  has not been compared
to micromechanical theories, and its general validity may,
therefore, be questioned. In complex flows (e.g.  flows characterized by shear localization, creep zones, and those influenced by boundaries~\cite{artoni18})
the rheology is known to become nonlocal, and the introduction of velocity 
fluctuations as an additional variable
seems a promising path. 

The strength of velocity fluctuations is a classical ingredient of diffusion theories. Dense kinetic theories for the segregation of binary mixtures of inelastic spheres \cite{jenkins89,arnarson98,arnarson04,larcher13,larcher15} predict diffusion coefficients that exhibit explicit dependence on {solid} fraction, the strength of the particle velocity fluctuations and the particles'{} size, mass, and collision properties. 
Consequently, our second  goal is to  characterize the scaling of self
diffusion with respect to granular temperature and compare the measured values of the self-diffusion coefficient to those predicted by dense kinetic theory.

%\paragraph{System parameters.}
\section{Discrete numerical simulations}
Simulations were performed by means of the open-source molecular dynamics software LAMMPS~\cite{lammps}.
A cubic simulation cell was used, with a size of 20x20x20 in particle diameters, in which the solid fraction $\phi$ was varied from 0.1 to 0.61. 
The results presented here are, however, restricted to the dense flow range of $\phi=$ 0.49 to 0.586 (number of particles $N=$ 7400 to 8838). 
The choice of the system size was motivated by the need of having a system larger than the typical correlation length (few particle diameters), but small enough to limit the computational cost. Previous literature \cite{oyama} suggested that $L=20d$ may be a good compromise. On our hand, we verified this by running dedicated simulations for a larger system size ($L=40d$, not shown here), which gave similar results in terms of self-diffusivity, and therefore support our choice for $L$. 
In order to avoid crystallization, a slight polydispersity was introduced:  the dimensionless diameter, $d$,  ranged uniformly from 0.9 to 1.1.
The mass of the particle with unit diameter was taken as the mass scale $m$. Then, in dimensionless terms,  the dimensionless mass density of the spheres is ${\rho}=6/\pi$. 

For the normal component of contact, a linear spring dashpot model, as employed by \citet{silbert01}, was used.  In this, the force between particles $i$ and $j$ is given by $F^{ij}_n = k_n \delta^{ij}_n -m^{ij}_{eff}\gamma_n \dot\delta^{ij}_n$, where $k_n$ is the normal stiffness, $\delta_n$ the particle interpenetration, $m^{ij}_{eff}={m_im_j}/{(m_i+m_j)}$ the effective mass of the interaction, and $\gamma_n$ the specific damping coefficient. For the tangential component, an elastic model with stiffness $k_t$,  no viscous damping and a frictional threshold was employed. The normal spring stiffness provides an intrinsic time scale for the system; $\sqrt{m/k_n}$ was used to nondimensionalize the time. The tangential stiffness was set as $k_t/k_n=2/7$, for the periods of normal and tangential oscillations to be equal \cite{schafer96}. The normal damping {coefficient, expressed in normalized units through the normalized time, was varied in the range $\gamma_n=(0.095-0.609)  \sqrt{k_n/m}$}, corresponding to a range of restitution coefficients  $e_{n}=0.5-0.9$ for the collision of two unit diameter particles. Note that, due to polydispersity, a slight heterogeneity of the effective restitution coefficient is expected. The coefficient of sliding friction was chosen to be ${\mu}$ = 0.5. {In molecular dynamics discrete element simulations, the computational time step $\Delta t$ is usually set as a  small fraction of the collision time, in order to ensure proper simulation of contact dynamics. The collision time for the linear spring-dashpot  model is  given by the relation: $t_{coll}=\pi \left[ {k_n}/{m_{eff}}-\left({\gamma_n}/{2}\right)^2\right]^{-1/2}$. In our case, the term containing the damping coefficient is quite negligible and therefore  $t_{coll}\approx {\pi}/{\sqrt{2}}\sqrt{{m}/{k_n}}$. The time step was therefore taken as $\Delta t=2\times10^{-2} \sqrt{m/k_n}$, which corresponds, approximately, to one-hundredth of a collision time.}

The numerical simulations were performed in simple shear under  fully periodic boundary conditions by means of the  LAMMPS `fix deform' scheme which, similar to the method used by \citet{radjai02}, applies shear by deformation of the simulation box. Note that this is different from \citet{campbell97} who used a ``sliding blocks'', Lees-Edwards \cite{lees} scheme for ensuring shear under periodic boundary conditions.
Coordinates $x$, $y$, and $z$ correspond in the directions of flow, vorticity, and gradient, respectively.  

Two types of simulations were carried out over the specified range of {solid} fraction: one with the shear rate constant over the range  $\dot \gamma = (10^{-6}-10^{-3}) \sqrt{k_n/m}$; the other with the pressure held approximately constant,  obtained by decreasing the shear rate while increasing the {solid} fraction. This was made to check that the results  depended on shear rate only trivially, as  is expected in the range of shear rates considered. Given that this was verified, we focus on  results obtained for a shear rate of  $\dot \gamma = 10^{-3} \sqrt{k_n/m}$.  

The simulation was carried out in {two} steps: first, the initial state was generated on a lattice partially filled with the desired distribution of particles, and {sheared  for $10^{9}$ steps} to ensure a steady state. {A Q6-analysis \cite{rintoul,richard99} was used to check that the shear applied during $10^9$ time steps was enough to remove any trace of the orientational order of the initial state.}  { In the second phase,  data analysis was performed, while the system was sheared at the same rate as in the preparation phase, again for  $10^{9}$ steps. In this phase,} snapshots of positions, velocities and inter-particle forces were extracted every 10$^{4}$ steps. {The cumulative deformation correponding to each of the two phases, preparation and analysis, was} therefore,  $\dot \gamma \Delta t =200$.  In contrast, the simulations of Radjai \& Roux \cite{radjai02} in two-dimensions had an applied total strain of about two, at an area fraction around 0.8.  The relatively large value of cumulative deformation employed here was necessary in order to determine the scaling of self-diffusion with time.  
The simulation output was treated in two steps: first, particle trajectories were reconstructed from time snapshots, removing the instantaneous mean field; then,  the statistics of the velocity probability distribution function and self-diffusion were calculated.

\section{Results}
\subsection{Velocity distribution functions} 

\begin{figure}[h]
\includegraphics[width=0.99\columnwidth]{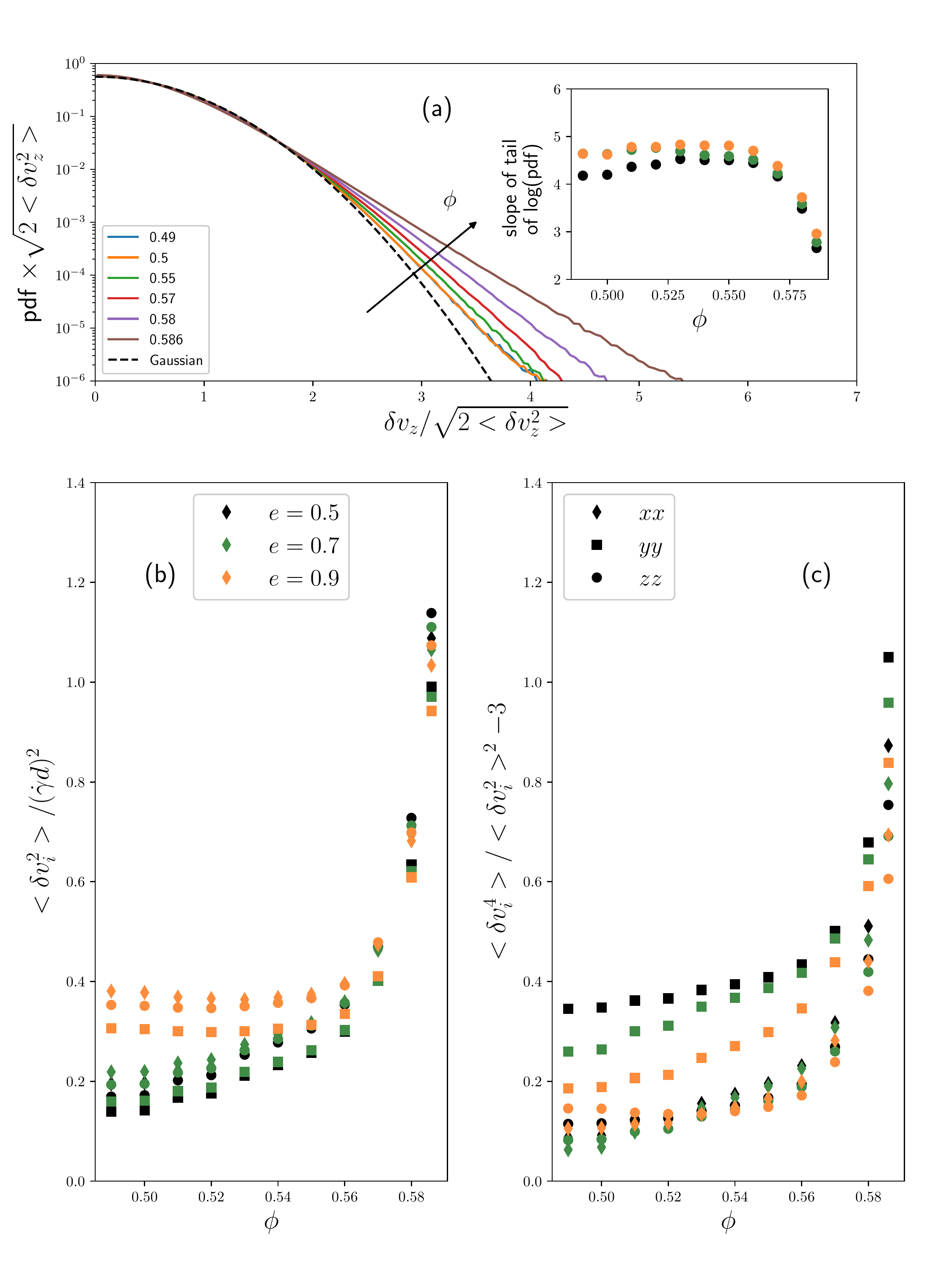}
\caption{(a) PDF of the instantaneous particle velocities for different values of the average solid fraction for $e_n=0.7$ (inset: slope of the logarithm of the high energy tail of the PDF as a function of the average solid fraction  for three values of the normal restitution coefficient, {colors for $e_n$ detailed in (b)}), (b) Dimensionless variance, and (c) excess kurtosis of the velocity distribution function for the three components of the velocity (symbols, detailed in (c)), and three values of the normal restitution coefficient (colors, detailed in (b)). 
}\label{fig:2}
\end{figure}

In Fig.~\ref{fig:2}a  we show the probability distribution function { (PDF) } of, for example,  the instantaneous $z-$velocity fluctuations for different values of the average solid fraction for $e_n=0.7$. It is evident that in dense systems the velocity distribution displays an exponential tail, as in the case of rate-independent shearing \cite{radjai02}, which broadens with increasing solid fraction. As the inset of Fig.~\ref{fig:2}a clarifies, such broadening of the exponential tail is important for $\phi>0.55$, and  quite independent of the restitution coefficient. 
Next, we describe the probability distribution function of each component. Note that for computing the velocity fluctuation in the $x-$direction,  we subtract the local mean velocity related to {the} mean shear ($\delta v_x = v_x - \dot \gamma (z-z_0)$, where $z_0$ is the center of the cell) \cite{artoni15}. In order to characterize the effect of $\phi$, $e_n$ and direction on the velocity PDF, we show two statistical descriptors for each component: the variance $\left\langle\delta v_i^2\right\rangle$ and the excess kurtosis $\left\langle\delta v_i^4\right\rangle/\left\langle\delta v_i^2\right\rangle^2-3$. The average of the variances in the three directions is the usual definition of the granular temperature, $T\equiv \sum_i\left\langle\delta v_i^2\right\rangle /3$. The excess kurtosis is a measure of the flatness of the distribution, and can be used to characterize the departure from a normal distribution, for which it is zero (it is equal to three for  a Laplace distribution). 

In Fig. \ref{fig:2}b, the variance of the different components of the velocity fluctuations behaves in a similar way, but the intensity of the fluctuations is different: for relatively low solid fractions, we observe $\left\langle\delta v_y^2\right\rangle    <\left\langle\delta v_z^2\right\rangle    < \left\langle\delta v_x^2\right\rangle    $; while, approaching $\phi=0.587$, the order is  $\left\langle\delta v_y^2\right\rangle    <\left\langle\delta v_x^2\right\rangle    < \left\langle\delta v_z^2\right\rangle    $.  Nevertheless, the anisotropy of velocity fluctuations appears to decrease with ${\phi}$.   With respect to the restitution coefficient,
for strong dissipation (points corresponding to $e_n\leq0.7$ in the figure), $\left\langle\delta v_i^2\right\rangle/(\dot\gamma d)^2$ is monotonically increasing with $\phi$, for the dense flows considered here. For larger values of the restitution coefficient (as in the case $e_n=0.9$ presented in the figure), the variance first decreases and then increases with $\phi$. Note that for dilute situations, $T/(\dot\gamma d)^2$ is reported to decrease with $\phi$ \cite{campbell97}. So, the value of the solid fraction for which the granular temperature displays a minimum seems to depend on $e_n$. Moreover, for low values of the solid fraction, the variances increase when increasing $e_n$; while, when approaching $\phi=0.587$, they appear to become independent of the restitution coefficient.

The excess kurtosis, shown in Fig.\ref{fig:2}c, generally increases with solid fraction, which mirrors the broadening of exponential tails with $\phi$ observed in Fig. \ref{fig:2}a. Deviation from a normal distribution appears to be stronger in the flow direction $x$. The effect of the restitution coefficient is also stronger for the flow direction: decreasing the coefficient of restitution  yields an increase of the excess kurtosis, which seems to be limited to low solid fractions. As was observed for the variance, the anisotropy between the velocity distributions is reduced when increasing $\phi$. Based on the data collected in Fig. \ref{fig:2}, we can conclude that  the fluctuation velocity vector distribution is non-Maxwellian in both the anisotropy and the exponential tail, and that the importance of these two effects depends on the values of the restitution coefficient and  the solid fraction.

\subsection{Self-diffusion}
\begin{figure}[h]
\includegraphics[width=0.9\columnwidth]{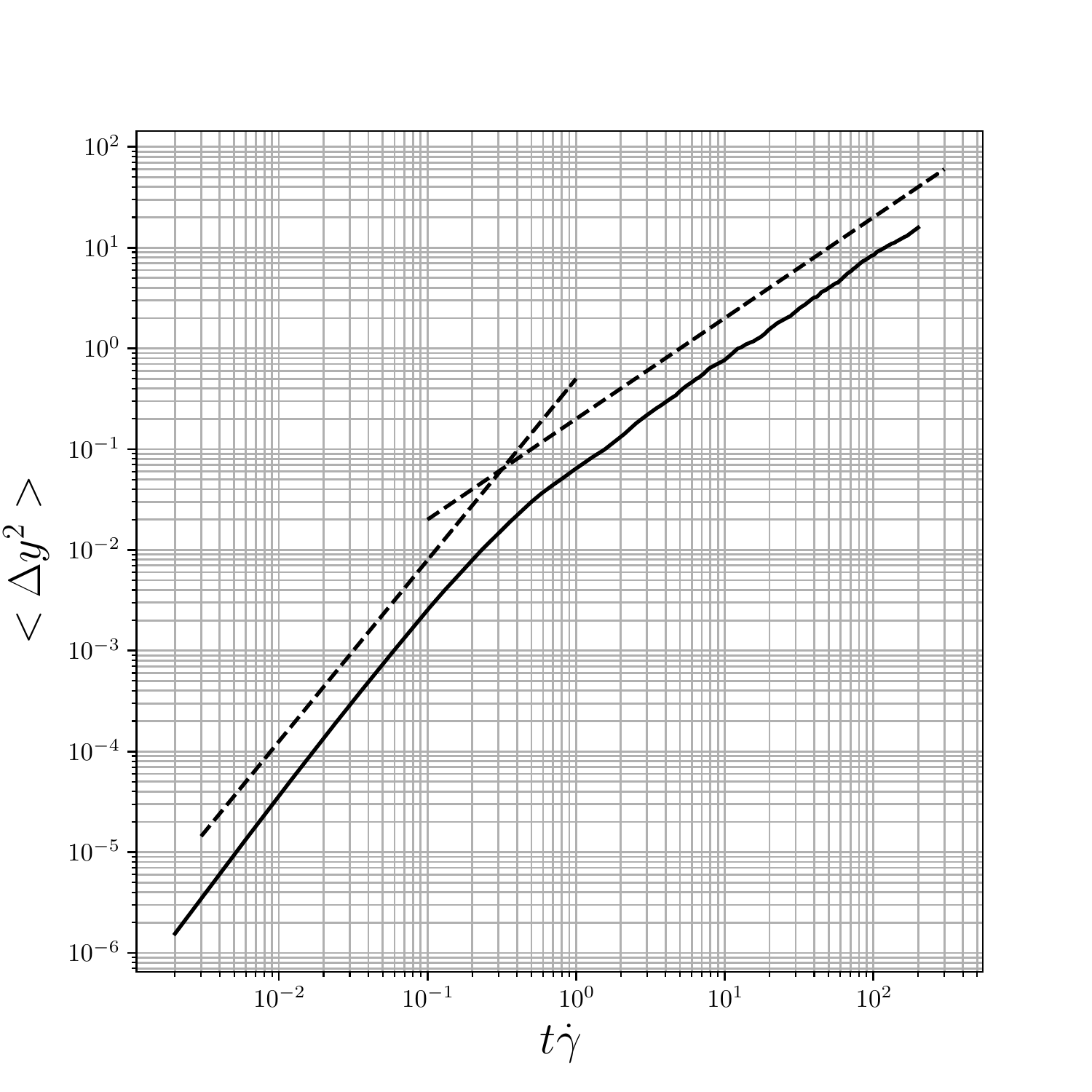}
\caption{ Average, cumulative squared displacement in the $y$ direction as a function of cumulative deformation, for $e_n=0.7$, $\phi=0.57$. The existence of two different power laws  is evident, one with exponent $\alpha\approx 1.8$ for low deformations, and one with exponent $\alpha\approx 1$ for large deformations. For large deformation the behavior is simple diffusive.  }\label{fig:3}
\end{figure}

The components of the diffusion tensor were determined by tracking the movement of the particles relative to their initial position, while taking into account the displacement due to the mean shear flow. We find that the particle self-diffusion, corresponding to correlation of displacements in directions $i$  and $j$, is proportional to a power of the time: 
\begin{equation}
%\left\langle \left(\Delta  x_i\right)^{2}\right\rangle \propto t^{\alpha },
\left\langle \Delta  x_i \Delta x_j\right\rangle \propto \Delta t^{\alpha },
\end{equation}
where the exponent is not constant.  As Fig. \ref{fig:3} exemplifies for the transverse $yy$-component, for small cumulative deformations ($\dot{\gamma }\Delta t<1$), an exponent $\alpha \approx 1.8$ is found, which corresponds to super-diffusive motions; while for large cumulative deformations ($\dot{\gamma }\Delta t >1$), a simple diffusive behavior is evident, with $\alpha \approx 1$. This double scaling is in agreement with previous results for dilute systems by Campbell \cite{campbell97}, and also with those in dilute collisional suspensions \cite{abbas09}. It seems to be the simple consequence, well known in turbulence \cite{taylor22, taylor53, taylor54a, taylor54b}, of the apparent diffusion associated with a random process with a finite correlation time (which is superdiffusive for short times). {For dense granular flows, particle displacements are constrained and frustrated by the mutual hindrance between neighboring particles, and therefore the physical mechanism behind their time evolution is different from turbulence. The present results show, however,  that the diffusive} behavior is present even in dense systems, and seem to indicate that the super-diffusive behavior observed by Radjai \& Roux \cite{radjai02} may be due the small cumulative deformation they employed.

\begin{figure}[h]
\includegraphics[width=\columnwidth]{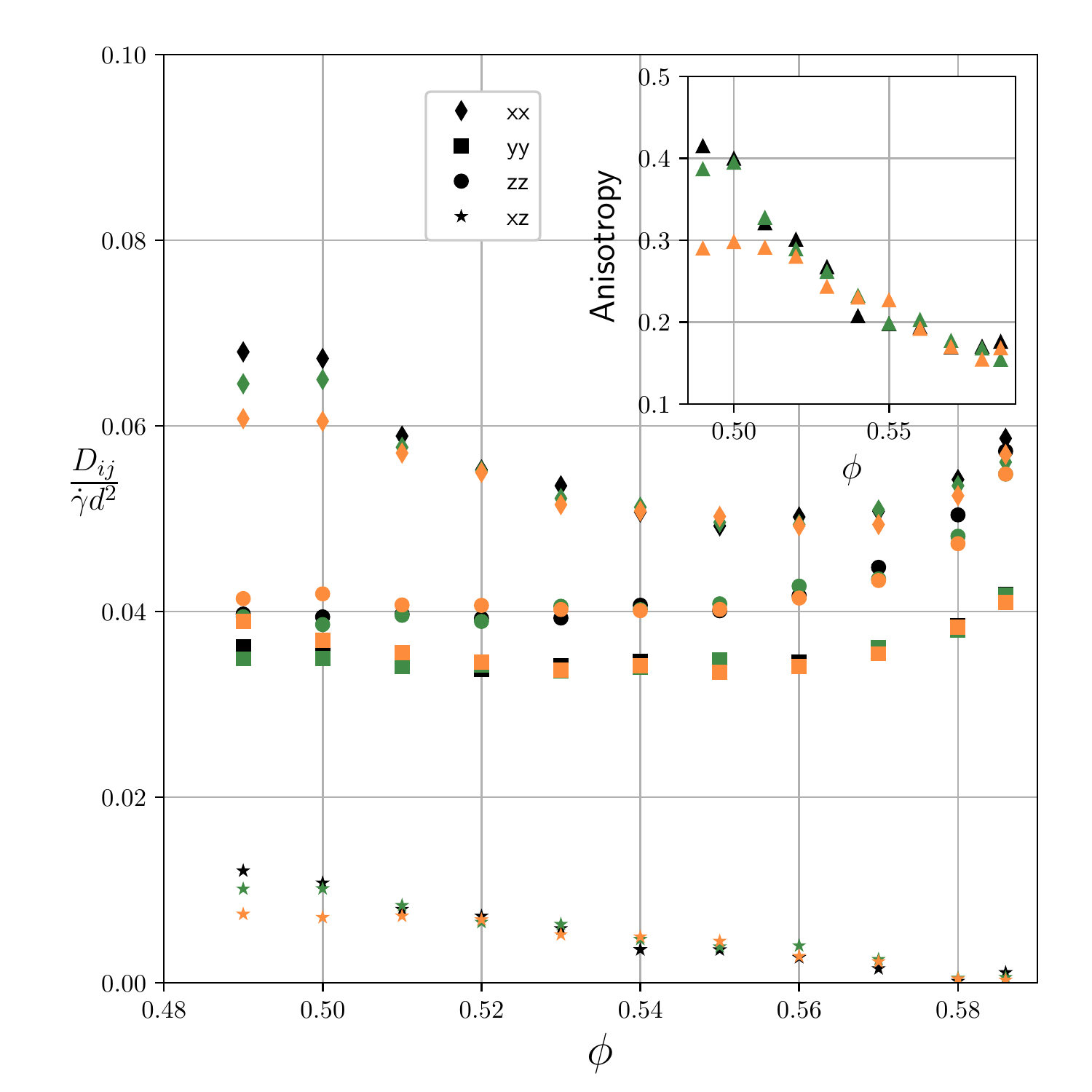}
\caption{The four non-zero components (symbols) of the self-diffusion tensor plotted together for different restitution coefficients (colors, same as in figure \ref{fig:2}) and different values of the solid fraction.  The tensor components are scaled by the product of the shear rate and squared particle diameter. The inset represents the anisotropy of the tensor as a function of the solid fraction via the measure detailed in the text.
}\label{fig1supp}
\end{figure}

\begin{figure}[h]
\includegraphics[width=1\columnwidth]{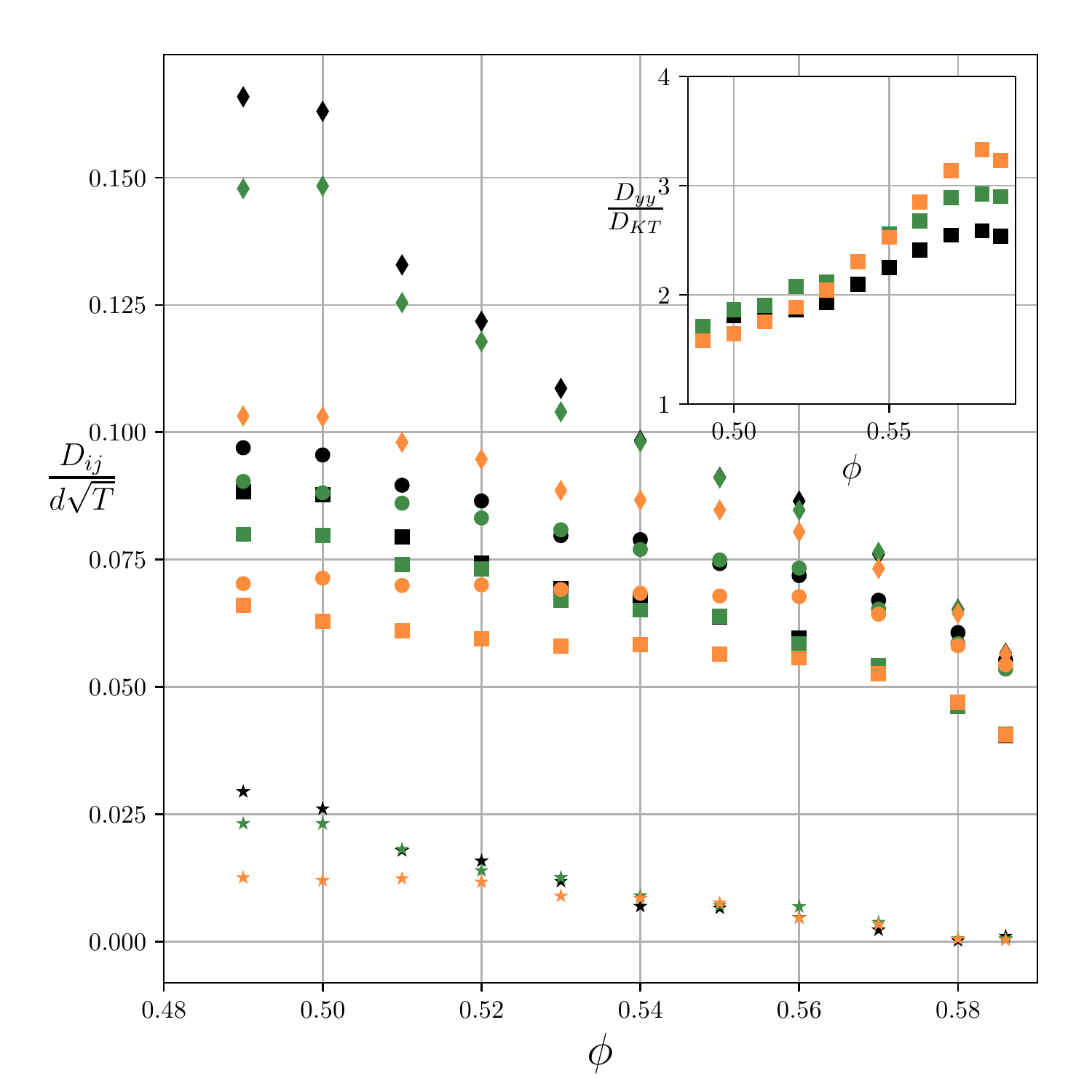}
\caption{The four non-zero components (symbols, same as in Fig.~\ref{fig1supp}) of the self-diffusion tensor plotted together for different restitution coefficients (colors, same as in Fig.~\ref{fig:2}) and different values of the solid fraction. The tensor components are scaled by the square root of the granular temperature.
The inset represents the ratio of the numerical $D_{yy}$ to the kinetic theory scaling, as a function of the solid fraction.}
\label{fig:4}
\end{figure}

Given that the behavior is diffusive for large cumulative deformations, it is possible to define a self-diffusion   tensor, as the limit for large cumulative deformations of the cumulative displacements correlations:
\begin{equation}
D_{ij}=\lim _{\Delta t\rightarrow \infty }\frac{\left\langle \Delta x_{i}\Delta x_{j}\right\rangle }{2\Delta t}.
\end{equation}
{For dilute granular shear flows ($\phi< 0.5$), \citet{campbell97} analyzed the components of the self-diffusivity tensor, scaled by $\dot \gamma d^2$, and showed  that the tensor was anisotropic with a clear hierarchy ($D_{xx} > D_{zz}  > D_{yy} > D_{xz}$),  and that the off-diagonal components other than \textit{xz} were negligible.}
In the following, we do not discuss the \textit{xy} and \textit{yz} components of the diffusion tensor, because, as in \citet{campbell97}, these components are negligible. The four other non-zero components are shown in {Figs. \ref{fig1supp} and \ref{fig:4}}. We first consider them normalized by the shear rate and the squared particle diameter, inspired by the empirical scaling previously proposed~\cite{utter04,fry19, cai19}. 

First, the off diagonal term seems to go to zero in the limit of $\phi\rightarrow \phi_c$.  Moreover, its magnitude is well below that of the diagonal terms. This appears to be a peculiarity of dense flows, $D_{xz}$ being comparable to $D_{yy}$ and $D_{zz}$ for dilute flows. 
On the other hand, regarding the dependence on solid fraction, it is evident that the diagonal components, scaled by the shear rate, display a nonmonotonic behavior, the strongest example being given by the streamwise, $xx$-component, which first decreases and then increases with $\phi$. 
Then, the diagonal terms display a moderate but evident anisotropy. 
Similarly to what is obseved in dilute flows \cite{campbell97}, we obtain $D_{xx} > D_{zz}  > D_{yy} > D_{xz}$. Yet the magnitude of the latter components are closer to each other in the dense case. 

The anisotropy of the self-diffusion tensor is  quantified by the indicator 
$$\sqrt{\frac{(D_1-D_2)^2 +(D_2-D_3)^2 + (D_3-D_1)^2 }{2 (D_1^2+D_2^3+D_3^2)}},$$ 
where the $D_i$ are the eigenvalues of the diffusion tensor, which is displayed in the inset of Fig.~\ref{fig1supp}. It is evident that anisotropy  decreases with increasing solid fraction, but does not disappear approaching $\phi_c$.  Finally, the diffusivities, scaled by the shear rate,  are independent of the restitution coefficient.
Based on these results, we can conclude that the empirical scaling  $D\approx 0.05 \dot\gamma d^2$ ~\cite{utter04,fry19, cai19}  gives the correct order of magnitude for the trace of the diffusivity tensor in the range of dense solid fractions considered.
Therefore, we think that the empirical scaling cited above may be employed in approximate analyses. Refined analyses must, however, take into account the nonmonotonic behavior  and anisotropy of the $D_{ij}$. 

Questions may be raised about the micromechanical origin of the scaling on diffusion on shear rate. A possible micromechanical framework to interpret this empirical result may be found in kinetic theory. 
As in isotropic turbulence~\cite{pope_2000}, expressions for the self-diffusivity in an isotropic dense granular gas can be represented by the formula $D=T \tau_v$ where $\tau_v$ is the time of autocorrelation of velocity fluctuations. In the kinetic theory of dense gases of elastic spheres {Chapman \& Cowling} \cite{chapman1990},  $\tau_v = {d \sqrt{\pi/T} }/ (16\phi g_0(\phi) )$, where $g_0(\phi)$ is the radial distribution function at contact.  From this the classical scaling {is obtained,} $D= {\sqrt{\pi T} d}/ (16\phi g_0(\phi)  )$, which was, for example, used by \citet{larcher13,larcher15} for a dense gas of frictional, slightly inelastic spheres.
The relevant parameter in such an expression that sets the time scale of diffusion is the strength of velocity fluctuations. 
Inelasticity in granular gases is known to increase the spatial and temporal span of velocity correlations, therefore increasing self-diffusion. In the isotropic case, this results in a correction to Chapman \& Cowling's formula, {and the scaling for the  self-diffusivity of the kinetic theory of granular gases  \cite{brilliantovbook} is therefore: \begin{equation} D_{KT}=\frac{\sqrt{\pi}}{8 (1+e)}\frac{d}{\phi g_0(\phi)} \sqrt{T}. \label{dkt}\end{equation} }

In order to evaluate this framework and the eventual modifications needed to adapt it to the dense systems considered here, we plot again in Fig.  \ref{fig:4} the self-diffusion tensor components, this time scaled by the square root of the granular temperature and by the particle diameter. 
As in kinetic theory, the ratios $D_{ij}/d\sqrt{T}$ are decreasing functions of $\phi$. Then, while the off diagonal component goes to zero as $\phi\rightarrow\phi_c$, the diagonal components remain finite. Moreover, for low values of the dense  solid fraction, the $D_{ij}/d\sqrt{T}$ depend on the restitution coefficient, which is not the case for high values of the dense solid fraction.

Given that the self-diffusivity tensor displays some anisotropy,  we compare isotropic scalings from kinetic theory to the transverse diffusivity component, $D_{yy}$, which is less affected by the anisotropy of shear. 
The radial distribution function at contact, $g_0(\phi)$, is often operationally defined through the equation of state for the pressure. The classical result {by Torquato}~\cite{torquato} for hard elastic frictionless spheres, for solid fractions  between freezing and random close packing  ({$\phi_F <\phi <\phi_{RCP}$, where $\phi_F=0.49$ and $\phi_{RCP}\approx0.64$}), is {\begin{equation}g_0(\phi)=5.6916\, \frac{\phi_{RCP} - \phi_F}{\phi_{RCP} - \phi}.\label{torq}\end{equation}} We note that in the framework of extended kinetic theory \cite{Jenkins2012, Berzi2014}, \citet{berzi15} have discussed constitutive relations  for frictional inelastic particles based on a contact radial distribution function that possesses a singularity at a critical solid fraction $\phi_c$ lower than the {random close packing} that depends on the friction coefficient. However, in our view, such singular behavior is appropriate for collisional transfers of momentum and energy, but not for those, such as diffusion, that involve transport of mass.

{Clearly, our numerical results support the choice of a radial distribution function not divergent at $\phi_c$, because the diagonal components remain finite when approaching the critical solid fraction. Therefore, {in the following, for estimating the theoretical self-diffusivity,  $D_{KT}$, we combine Eq. \ref{dkt} with  the classical radial distribution function by \citet{torquato}, Eq. \ref{torq}. } In the inset of Fig. \ref{fig:4}, we plot the ratio between the numerically obtained $D_{yy}$ and the prediction from kinetic theory, $D_{KT}$, by employing the classical radial distribution function by \citet{torquato}. It is evident that, for low solid fractions, the kinetic theory prediction is not far from the measured values, and correctly models the effect of the restitution coefficient. However, for denser systems, the deviation increases and the ratio $D_{yy}/D_{KT}$ reaches a value of about three.} 
{Note that the deviations from the kinetic theory observed for dense systems have also been reported experimentally and numerically in 2D systems of disks submitted to random fluctuations induced by an air table~\cite{oger96}.}
 In the following subsection, we discuss the possible origin for the differences between kinetic theory and the simulation.

\subsection{Correlated bulk motion}In order to determine whether part of the deviation from kinetic theory could come from correlated bulk motions, we computed energy spectra of spatial velocity fluctuations. For each time snapshot of the system, we first interpolated the particle velocity fluctuations on a regular grid, then obtained two-point velocity correlations and the energy spectrum through the (spatial) Fourier transform of the interpolated field. As in turbulence \cite{pope_2000, oyama}, the energy spectrum was spherically averaged with respect to the wavenumber $k$.

\begin{figure}[h]
\includegraphics[width=1\columnwidth]{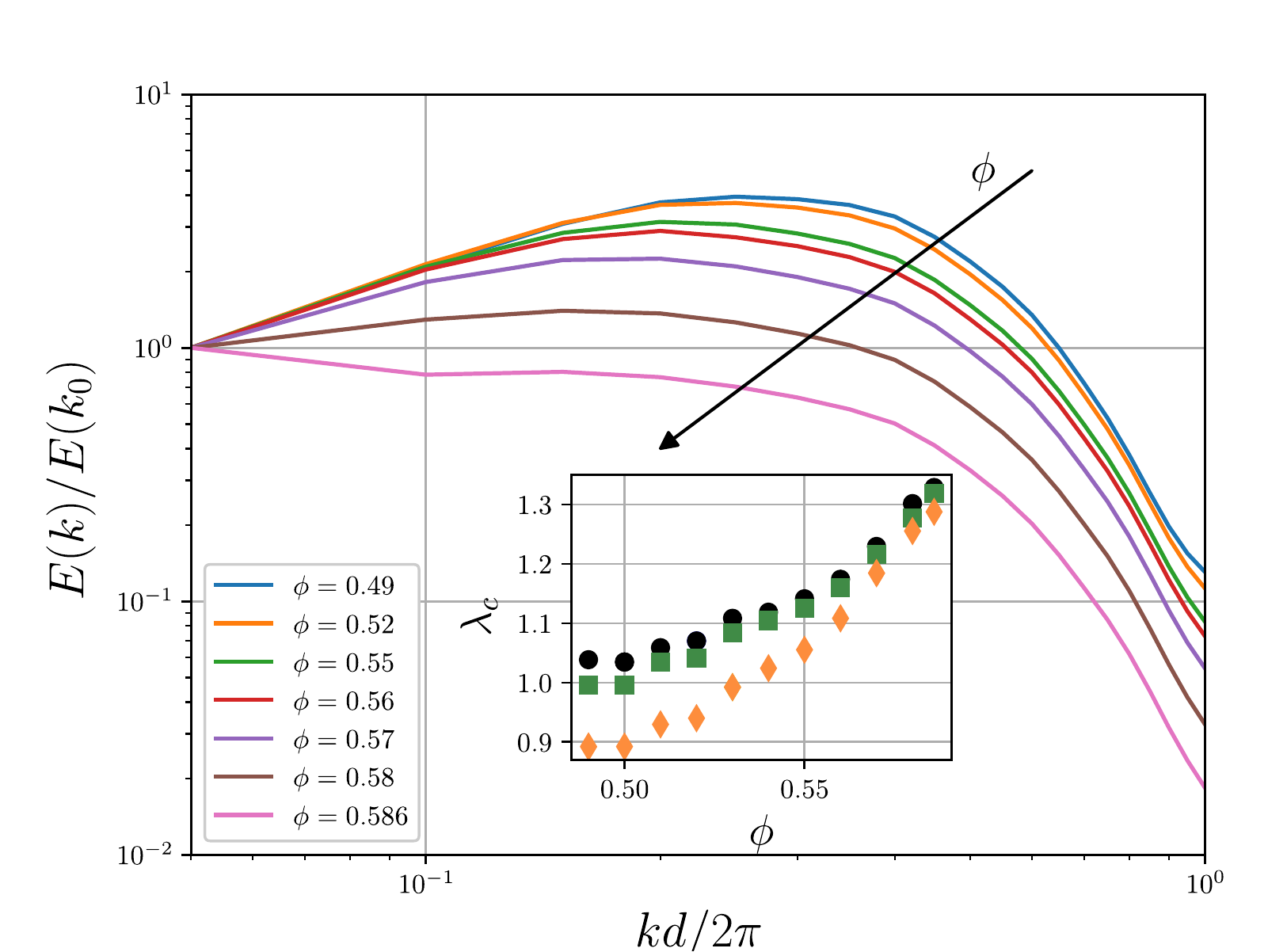}
\caption{ Power spectra for $e_n=0.7$ and  several solid fractions. The inset represents the integral length scale obtained from two-point correlations, as a function of the solid fraction, for different restitution coefficients (colors, same as in Fig.~\ref{fig:2}). }
\label{fig:5}
\end{figure}

In Fig. \ref{fig:5} we plot the energy spectra, normalized by the value at the smallest wavenumber, {$k=2 \pi / L$}. It is evident that the spectral energy density is not a monotonic function of the wavenumber: particularly for lower dense solid fractions, the spectra display a maximum and then decrease with, ultimately, a power-law cutoff. The energy-containing scale represented by the position of the maximum of the spectrum, slightly depends on solid fraction, as does the integral length scale extracted from the two-point correlation functions, varying between one and two particle diameters. It is evident that velocity  correlations associated with several particles exist and contribute energy to the spectra.  

In granular shearing flows, there is not the same separation of scales as in molecular
gases. However, in dense granular shearing flows, the particles interact over a length scale that is
the distance between their edges, which is a fraction of their diameter. {At this scale, pre-collisional velocity correlations exist as shown by  \citet{mitarai}, which affect the amount of velocity fluctuations.} With this in mind, we
suggest that a more appropriate measure of the temperature may be the energy of the velocity
fluctuations at a scale below the particle diameter, {in relation to pre-collisional velocity correlations}. Velocity fluctuations evaluated at lengthscales of the order of tens of
particle diameters, then, may be analogous to the macroscopic fluid velocity fluctuations that
contribute to a turbulent eddy diffusivity. Although in the present soft particle simulation {it was not possible to implement a detailed follow-up of collisions, and therefore}  it is
impossible to access length scales smaller than the diameter of the spheres, we note that the
energy of the velocity fluctuations at this scale is an order of magnitude less than that at the
largest length scales. Hard sphere simulations of the type carried out by \citet{mitarai}, in a study of pre-collisional velocity correlations, seem to permit access to the velocity fluctuations at smaller length scales.

\section{Conclusions} 
We have characterized the properties of the velocity distribution function and the components of
the self-diffusion tensor in discrete numerical simulations of dense, steady, homogeneous
shearing flows of nearly identical, inelastic, frictional spheres. 
We have discussed the anisotropy of the self-diffusion tensor, and its dependence on solid fraction. Our results provided a test for the empirical scaling $D\sim 0.05  d^2 \dot\gamma$ in  a wide range of dense solid fractions. We found that, although such a scaling gives the correct order for the diagonal components of the self-diffusivity, a nonmonotonic dependence on $\phi$ as well as a moderate anisotropy are present, which may be important in refined analyses.

In order to look for micromechanical explanations for the scaling cited above, we
 compared the values for one of
the components of the self-diffusion to that predicted by kinetic theory. When the strength of the
velocity fluctuations at all length scales was employed as the granular temperature, kinetic theory
was found to under-predict self-diffusion. A spectral analysis of the velocity fluctuations
indicated how the strength of the velocity fluctuations varied with their wavelength. This
variation led us to suggest that there may be an effective separation of scales between
fluctuations at a fraction of diameter, at which the particles interact, and those larger than a
particle diameter that may be the analog of the eddies of a turbulent fluid \cite{miller13, griffani13}. 
This effective separation of scales might be at the origin of the deviation from kinetic theory. 

Our aim is to obtain relations applicable to the evolution in space and time of granular segregation in industrial processes and geophysical flows, in which granular diffusion takes place at high {solid} fractions. It is well known that in such heterogeneous flows, granular temperature is an important dynamic variable, which can be used to model nonlocal effects. Here we have shown that it is possible to model the magnitude of the self-diffusion tensor by an empirical law based on shear rate~\cite{utter04,fry19, cai19}, but also by kinetic theory, provided that a correction for dense systems is introduced.
{This correction can be written as:
\begin{equation}
D^*_{KT} = \xi(\phi)  \frac{\sqrt\pi}{8(1+e)} \frac{d}{\phi g_0(\phi)} \sqrt{T},
\end{equation}}
where $\xi$ is a correlation factor corresponding to the inset of Fig.~\ref{fig:4}. It is important to consider such a scaling based on kinetic theory partly because of its micromechanical origin, and partly because at present the effect of nonlocality on self-diffusion is not clear. Further research will deal with the measurement of self-diffusivities in heterogeneous flows in order to {provide more evidences concerning the correcting factor $\xi$ and} determine the relative validity of the two frameworks discussed above.

\section*{Conflicts of interest}There are no conflicts to declare.

\section*{Acknowledgements}
The numerical simulations were carried out at the CCIPL (Centre de Calcul Intensif des Pays de la Loire) under the project ``Simulation num\'erique discr\`ete de la fracture des mat\'eriaux granulaires".

%%%END OF MAIN TEXT%%%

%The \balance command can be used to balance the columns on the final page if desired. It should be placed anywhere within the first column of the last page.

\balance

%If notes are included in your references you can change the title from 'References' to 'Notes and references' using the following command:
%\renewcommand\refname{Notes and references}

%%%REFERENCES%%%
%\bibliography{rsc} %You need to replace "rsc" on this line with the name of your .bib file
%\bibliographystyle{rsc} %the RSC's .bst file
\providecommand*{\mcitethebibliography}{\thebibliography}
\csname @ifundefined\endcsname{endmcitethebibliography}
{\let\endmcitethebibliography\endthebibliography}{}

\end{document}